\documentclass[twocolumn,astrosym]{aastex62}
\DeclareSymbolFont{cmletters}{OML}{cmm}{m}{it}
\DeclareMathSymbol{v}{\mathalpha}{cmletters}{"76}

\graphicspath{{./}{figures/}}

\received{November 21, 2022}
\revised{January 18, 2022}
\accepted{January 30, 2022}
\submitjournal{ApJL}

\shorttitle{Collapsars from Real Stars}
\shortauthors{Halevi et al.}

\begin{document}

\title{Density Profiles of Collapsed Rotating Massive Stars Favor Long Gamma-Ray Bursts}

\correspondingauthor{Goni Halevi}
\email{ghalevi@princeton.edu}

\author[0000-0002-7232-101X]{Goni Halevi}
\affiliation{Department of Astrophysical Sciences, Princeton University, 
4 Ivy Lane, Princeton, NJ 08544, USA}
\affiliation{Institute For Advanced Study, 1 Einstein Drive,
Princeton, NJ 08540, USA}

\author{Belinda Wu}
\affiliation{Department of Astrophysical Sciences, Princeton University, 
4 Ivy Lane, Princeton, NJ 08544, USA}

\author[0000-0002-9371-1447]{Philipp Mösta}
\affiliation{GRAPPA, 
Anton Pannekoek Institute for Astronomy, Institute of
High-Energy Physics, and Insitute of Theoretical Physics, University of Amsterdam,
Science Park 904, 1098 XH Amsterdam, The Netherlands}

\author[0000-0003-3115-2456]{Ore Gottlieb}
\affiliation{Center for Interdisciplinary Exploration \& Research in Astrophysics (CIERA), Physics \& Astronomy, Northwestern University, Evanston, IL 60201, USA}

\author[0000-0002-9182-2047]{Alexander Tchekhovskoy}
\affiliation{Center for Interdisciplinary Exploration \& Research in Astrophysics (CIERA), Physics \& Astronomy, Northwestern University, Evanston, IL 60201, USA}

\author[0000-0002-3874-2769]{David R. Aguilera-Dena}
\affiliation{Institute of Astrophysics, FORTH, Dept. of Physics, 
University of Crete, Voutes, University Campus, GR-71003 Heraklion, 
Greece}

\begin{abstract}

Long-duration gamma-ray bursts (lGRBs) originate in relativistic collimated outflows -- jets -- that drill their way out of collapsing massive stars. Accurately modeling this process requires realistic stellar profiles for the jets to propagate through and break out of. Most previous studies have used simple power laws or pre-collapse models for massive stars. However, the relevant stellar profile for lGRB models is in fact that of a star after its core has collapsed to form a compact object. To self-consistently compute such a stellar profile, we use the open-source code GR1D to simulate the core-collapse process for a suite of low-metallicity, rotating, massive stellar progenitors that have undergone chemically homogeneous evolution. Our models span a range of zero-age main sequence (ZAMS) masses: $M_\mathrm{ZAMS} = 13, 18, 21, 25, 35, 40$, and $45 M_\sun$. All of these models, at the onset of core-collapse, feature steep density profiles, $\rho \propto r^{-\alpha}$ with $\alpha\approx 2.5$, which would result in jets that are inconsistent with lGRB observables. We follow the collapse of four out of our seven models until they form BHs and the other three proto-neutron stars (PNSs). We find, across all models, that the density profile outside of the newly-formed BH or PNS is well-represented by a flatter power law with $\alpha \approx 1.35{-}1.55$. Such flat density profiles are conducive to successful formation and breakout of BH-powered jets and, in fact, required to reproduce observable properties of lGRBs. Future models of lGRBs should be initialized with shallower \textit{post-collapse} stellar profiles like those presented here instead of the much steeper pre-collapse profiles that are typically used.

\end{abstract}

\keywords{hydrodynamical simulations --- 
stellar mass black holes --- gamma-ray bursts --- core-collapse supernovae}

\section{Introduction} \label{sec:intro}
Evolved, massive stars that undergo core-collapse are commonly accepted as the progenitors of luminous, energetic transient sources including long-duration gamma-ray bursts (lGRBs) and core-collapse supernova (CCSN) explosions. The environments of lGRBs  -- star-forming, preferentially low-metallicity galaxies -- suggest a link to the core-collapse process \citep[e.g.][]{bloom02}. Individual lGRBs detected to be coincident with Type Ic-bl SNe \citep[e.g.][]{modjaz06,galama98}, hydrogen- and helium-deficient explosions feature broad spectral features, bolster the `SN-GRB' connection between the central engines driving these events \citep{modjaz16}. One of the favored models for this engine is the collapsar scenario \citep{macfadyen99}. Under this framework, the iron core of a massive star collapses to a Kerr black hole (BH) without driving a successful SN explosion. Afterwards, the high angular momentum of the stellar envelope leads to disk formation and the BH accretes matter, enabling energy to be extracted as Poynting flux \citep{blandford77,komissarov09}. This process leads to the launching of a relativistic, collimated outflow -- a jet -- that clears its path out of the star and breaks out of the envelope to eventually be observed as a burst of beamed, energetic gamma rays. 

Evolved, massive stripped-envelope stars, such as Wolf-Rayet (WR) stars, are theoretically the most likely evolutionary channel to lead to collapsars as their significant wind-driven mass loss results in depleted stellar envelopes for jets to break out of \citep{woosley93}. Stripped envelope stars are also associated with SNe that lack strong hydrogen or helium features in their spectra -- Type Ic SNe \citep[e.g.][]{2014Natur.509..471G,2017A&A...603A..51D}. 
This same class of SNe is observationally linked to lGRBs, suggesting that WRs or other stripped-envelope stars are the progenitors of lGRBs. These stars may include ones that have undergone chemically homogeneous evolution (CHE) without developing optically thick winds characteristic of WRs, which are observationally distinguished by their emission lines \citep[e.g.][]{2015A&A...581A..15S, 2022A&A...661A..60A}. Massive, stripped-envelope stars that fail to form BHs may also successfully power lGRBs and SNe Ic-BLs through the proto-magnetar model \citep{protomagnetar_GRB,protomagnetar_IcBL, 2023arXiv230105401S}. Simulations suggested that the central engine of a lGRB can itself trigger a SN Ic-BL in a WR star, jointly producing both observed phenomena from the same massive stellar progenitor \citep{2018ApJ...860...38B}, but others have found that the quasi-spherical explosion and beamed relativistic emission instead require two distinct energy channels \citep{2022MNRAS.517..582E}.

A subset of SNe Ic known as Type I superluminous supernovae (SLSNe) have also been proposed to have similar origins to lGRBs \citep[e.g.][]{2014ApJ...787..138L,2016MNRAS.458...84A,2018ApJ...858..115A,2018MNRAS.475.2659M} based on both theoretical and observational evidence. SLSNe exhibit intrinsic luminosities one to two orders of magnitude greater than those of ``normal'' SNe \citep[see][for reviews]{2012Sci...337..927G, 2018SSRv..214...59M}. Their spectral signatures point to progenitors that have undergone severe mass loss and/or mixing phases \citep{2016MNRAS.458.3455M,2019ApJ...882..102G}, while their extreme luminosities suggest a source of power in addition to the radioactive decay of $^{56}$Ni that powers other SNe. Stars that have undergone CHE that fail to form BHs but instead leave behind millisecond magnetars are proposed to power these SLSNe \citep[SLSNe;][]{2010ApJ...717..245K,2010ApJ...719L.204W,2015MNRAS.454.3311M,2017ApJ...850...55N} by continuously depositing energy into the ejecta during spin-down. Self-consistent simulations of both SLSNe and lGRBs require realistic stellar models like those presented in this Letter.

Numerical studies of lGRBs \citep[see][for a review]{2015JHEAp...7...17L} often manually inject a jet near the center of a stellar profile and then follow its evolution to calculate observables like light curves and spectral signatures. Alternatively, simulations can capture the ab-initio engine formation but are then computationally limited to following the jet for too short a duration to extract observable properties \citep[e.g.][]{burrows07,mosta14,halevi18,2020MNRAS.492.4613O}. Many of these numerical studies \citep[e.g][]{2007ApJ...665..569M,lazzati12,2013ApJ...767...19L,2016ApJ...826..180L,2018ApJ...860...38B,2019ApJ...880..135X} use pre-collapse WR stellar models as their lGRB progenitors, most commonly the 16TI model of \citet{2006ApJ...637..914W}. The choice of these stellar models assumes that the process that leads to BH or protomagnetar formation leaves the rest of the star unaffected.

Recent three-dimensional (3D) general-relativistic magnetohydrodynamical (GRMHD) simulations \citep{gottlieb22a,gottlieb22b} are the first to self-consistently launch a jet through accretion onto a Kerr BH and follow its journey through the stellar envelope, spanning the large spatial and temporal range necessary to study the observable jet properties. By exploring a range of stellar density profiles represented by analytic power-laws rather than simply assuming a fit to 16TI, these simulations showed that the steepness of the density profile constrains the physical properties of the jet \citep{gottlieb22a}. In particular, \citet{gottlieb22a} find that only those profiles that are significantly shallower than 16TI and other similar pre-collapse WR stars are compatible with observations of lGRBs.

In this Letter, we investigate the core-collapse process in massive stars stripped-envelope stars, which are the likely progenitors of lGRBs, and their post-collapse properties. In \S \ref{sec:methods}, we describe the properties of the stellar models from MESA that we evolve and our numerical setup for doing so. We report the results from our core-collapse GR1D simulations, including the nature of the compact remnants and density profiles in \S \ref{sec:results}. We end by discussing the consequences of these results in the context of lGRBs and encouraging future models to use physically-motivated post-core-collapse stellar profiles in \S \ref{sec:conclusion}.

\section{Methods} \label{sec:methods}

In this work, we present results from simulations of the core-collapse of massive stars and investigate their evolution and final, post-collapse states. 

\subsection{Pre-collapse stellar models}
The initial conditions for our core-collapse simulations are drawn from the stellar evolution models described in \citet[][hereafter AD20]{aguileradena20}.

AD20 presents the evolution, until the onset of core-collapse, of stellar progenitors for massive, low-metallicity stars, spanning zero-age main sequence (ZAMS) masses of $M_\mathrm{ZAMS} = (4-45)M_\sun$. The models are computed using the open-source 1D stellar evolution Modules for Experiments in Stellar Astrophysics, version 10398 \citep[MESA;][]{Paxton2011,Paxton2013,Paxton2015,Paxton2018}. Each star is initialized with (1) a rapid equatorial rotational velocity of $600~\mathrm{km}~\mathrm{s}^{-1}$ and (2) a low metallicity of (1/50)$Z_{\sun}$, where $Z_\sun$ represents the solar metallicity with abundances scaled from \citet{grevesse96}. The fast initial rotation guarantees effective mixing leading to quasi-CHE. These stellar models, which result in fast-rotating pre-collapse cores with hydrogen- and helium-depleted envelopes, are proposed as potential progenitors of lGRBs, and SLSNe \citep[][AD20]{2018ApJ...858..115A} or SNe Ic-BL.

We evolve seven different models from the set of 42 included in AD20, spanning a range of initial masses and expected explosion properties. We choose the $M_\mathrm{ZAMS} = 18 M_\sun$ model as our fiducial model because it is closest in mass, at the onset of core-collapse, to the well-studied lGRB progenitor model 16TI of \citet{woosley06}. By the start of core-collapse, this model has a mass of 14.15$M_\sun$ (compared to 13.95$M_\sun$ for 16TI). We note that the mass loss in the the AD20 suite of models is dictated by rotation and enhanced by neutrino-driven contraction, while the 16TI model loses mass due to WR winds. The true nature and rate of mass loss in such low metallicity, rapidly rotating stars is debated, as their winds may not be sufficiently optically thick to qualify as WR stars \citep[e.g.][]{2015A&A...581A..15S,2020A&A...634A..79S}. 

The $M_\mathrm{ZAMS} = 18M_\sun$ stellar model is classified by AD20 as a likely failed SN and potential lGRB progenitor based on various explosion criteria. The simplest of these is the core-compactness parameter, $\xi_M$, which is motivated by hydrodynamic simulations of neutrino-driven SNe and defined as
\begin{equation}
    \xi_M = \frac{M/M_\sun}{R(M)/1000~\mathrm{km}}, \label{eqn:compactness}
\end{equation}
where $R(M)$ is the radius of the enclosed baryonic mass $M$ \citep{2011ApJ...730...70O}. This single-parameter estimate is commonly used as an indicator of `explodability'-- whether or not a non-rotating stellar core will lead to a successful neutrino-driven explosion. It is often measured at a mass coordinate of 2.5 $M_\sun$, which corresponds to a typical infall velocity of 1000 km s$^{-1}$. Non-rotating cores with $\xi_{2.5} \gtrsim 0.45$ are predicted to be difficult to successfully explode, as calibrated by core-collapse simulations  \citep{sukhbold2014}. Our fiducial model has $\xi_{2.5} = 0.62$. All of the models we present here, except for the least massive ($M_\mathrm{ZAMS} = 13M_\sun$), have compactness parameters above 0.45 and are therefore expected by this simple explodability predictor to fail to explode and to instead collapse to form BHs.

\begin{figure}
    \centering
    \includegraphics[width=0.9\linewidth]{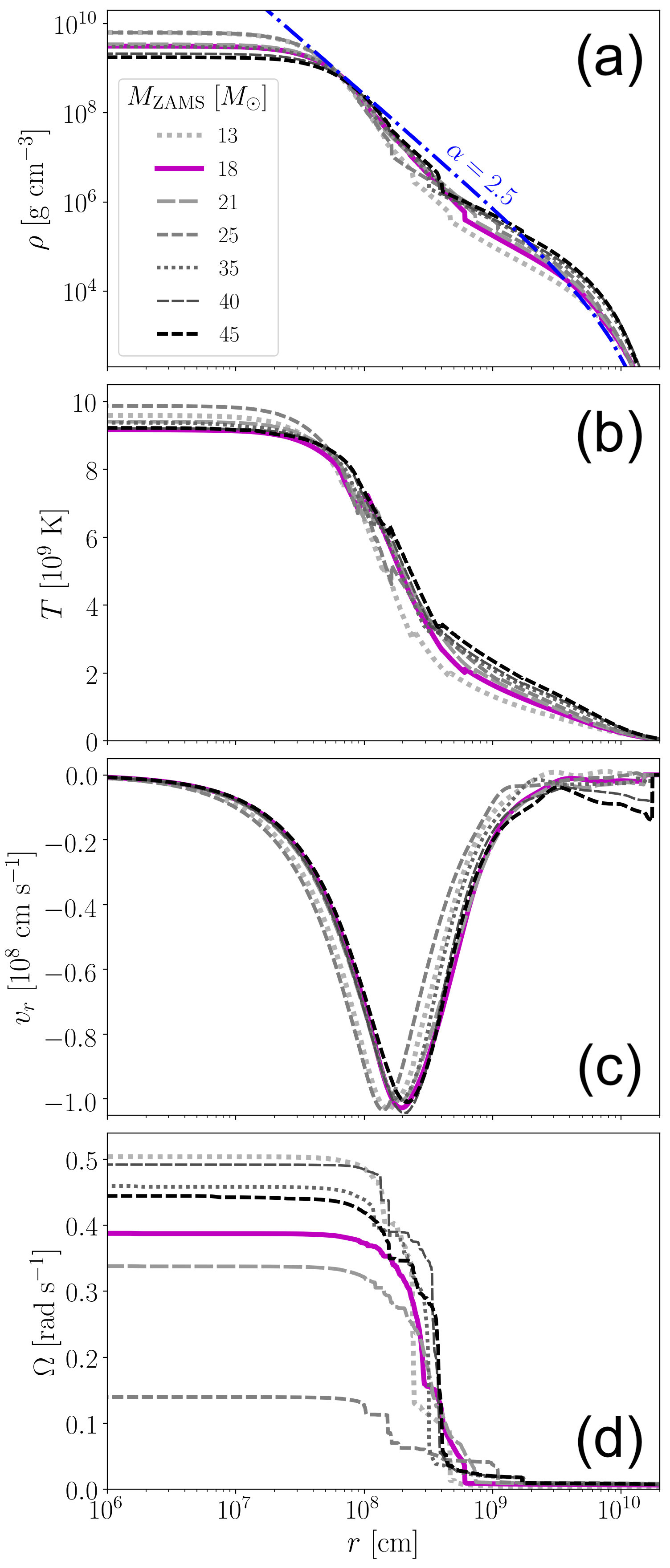}
    \caption{Stellar properties as functions of radius for the MESA models of \citet{aguileradena20} at the onset of core-collapse. In the four panels, we show (a) mass density, (b) temperature, (c) radial velocity, and (d) angular velocity. The different mass models are represented by different color and style lines as labeled, with the fiducial $M_\mathrm{ZAMS} = 18M_\sun$ model in solid magneta. We also include, in blue dash-dotted line, a fit to Equation (\ref{eqn:density}) with $\alpha = 2.5$, which represents the pre-collapse density profile characteristic of massive stellar progenitors that is ruled out by observed properties of lGRBs \citep{gottlieb22a}. All models share similar density, temperature, and radial velocity profiles. Their core rotational velocities differ by factors of a few and vary non-monotonically with ZAMS mass.}
    \label{fig:initialstellarprofs}
\end{figure}

Figure \ref{fig:initialstellarprofs} shows radial profiles of density, temperature, radial velocity, and angular velocity at the onset of core-collapse for the fiducial $M_\mathrm{ZAMS} = 18M_\sun$ model along with the other six models. We include the initial parameters for each of these models, as taken from AD20, in the first few columns of Table \ref{tab:modelprops}. In particular, we list the stellar mass at the onset of core-collapse (defined as the time when the core infall velocities first exceed 1000 km s$^{-1}$) $M_\mathrm{pre-cc}$ and the compactness parameter $\xi_{2.5}$. We note that while there are many similarities between the different models, there are also non-linear differences between them, for example in the rotational velocities of their core regions (see Figure \ref{fig:initialstellarprofs}d). This is also reflected in the non-monotonic (as a function of stellar mass) behavior of $\xi_{2.5}$.

Besides the $M_\mathrm{ZAMS} = 13M_\sun$ model, all of the models we adopt fail to meet the explosion criterion of \citet{muller2016}, which predicts properties of neutrino-driven explosions based on a semi-analytic model for stellar structure. Both the least ($13M_\sun$) and most massive ($45M_\sun$) models are predicted to explode based on the \citet{ertl2016} test, which employs a two-parameter representation of stellar structure, while the rest fail to meet this criterion as well. These predictions are also not monotonic with initial or pre-collapse mass due to the complexities of stellar evolution. We include the $M_\mathrm{ZAMS} = 13M_\sun$ model as a comparison point to the failed SNe, but from the perspective of likely lGRB progenitors, we focus on the more massive models.

\begin{table}
    \centering
    \begin{tabular}{c|cc|cclc}
    \hline
    $M_\mathrm{ZAMS}$ & $M_\mathrm{pre-cc}$ & $\xi_{2.5}$ & $t_\mathrm{bounce}$ & $t_f$ & $M_\mathrm{core}$ & $\alpha$ \\ 
    ($M_\sun$) & ($M_\sun$) &  & (ms) & (ms) & ($M_\sun$) & \\ \hline
    13 & 10.37 & 0.21 & 152 & 835 & 2.03 & 1.42 \\
    18 & 14.15 & 0.62 & 234 & 741 & 2.45$^*$ & 1.45 \\
    21 & 16.39 & 0.66 & 216 & 719 & 2.44$^*$ & 1.43 \\
    25 & 19.33 & 0.47 & 152 & 767 & 2.08 & 1.37 \\
    35 & 26.53 & 0.57 & 247 & 825 & 2.40 & 1.36 \\
    40 & 30.08 & 0.78 & 258 & 735 & 2.68$^*$ & 1.55 \\ 
    45 & 33.59 & 0.85 & 285 & 614 & 2.63$^*$ & 1.44 \\
    \hline
    \end{tabular}

    \caption{Parameters of the stellar models. We take the following pre-collapse quantities directly from AD20: ZAMS mass $M_\mathrm{ZAMS}$, mass at the time of re-mapping to GR1D $M_\mathrm{pre-cc}$, and compactness parameter at this time $\xi_{2.5}$ as defined in Equation (\ref{eqn:compactness}). We add to these the time of core-bounce in our GR1D simulations $t_\mathrm{bounce}$ and final quantities at the end of the simulation, which occurs at $t_f$: the mass of the inner core $M_\mathrm{core}$ (with an asterisk indicating BH-formation), and the best-fit power-law index for the density profile $\alpha$.}
    \label{tab:modelprops}
\end{table}

\subsection{Core-collapse simulations}
As we discuss below, we map each of the pre-collapse models onto GR1D \citep{oconnor10}, a spherically-symmetric, general-relativistic neutrino hydrodynamics code, in order to evolve through the final stage of stellar evolution. GR1D is an open-source tool\footnote{\url{https://github.com/evanoconnor/GR1D}} for simulating core-collapse and BH formation. It uses a finite-volume scheme with piecewise-parabolic reconstruction and a Riemann solver to solve the discretized equations of general-relativistic hydrodynamics (GRHD). It couples with microphysical, tabulated equations of state and includes an effective rotation that makes it effectively 1.5D. It too is a modular code and, crucially, implements neutrino transport in the M1 formulation \citep{oconnor15} with tabulated multi-group neutrino opacities. Spherically symmetric models can never fully capture the inherently multi-dimensional properties of stars (e.g. critically-rotating stars are expected to be significantly oblate) and their collapse process \citep[e.g.][]{2016PASA...33...48M,2018ApJ...865...81O}. However, for the purpose of approximating a collapsing star's remnant mass, and averaged radial thermodynamic profiles rather, GR1D is a useful and sufficient tool \citep[e.g.][]{oconnor10, 2011ApJ...730...70O,2018JPhG...45j4001O}. GR1D allows us to simulate the entire star rather than collapsing only the inner-region, which is critical for understanding the environment that a collimated outflow from the accreting newborn compact object must propagate through and break out of in order to power a lGRB. 

\begin{figure*}
    \centering
    \includegraphics[width=\linewidth]{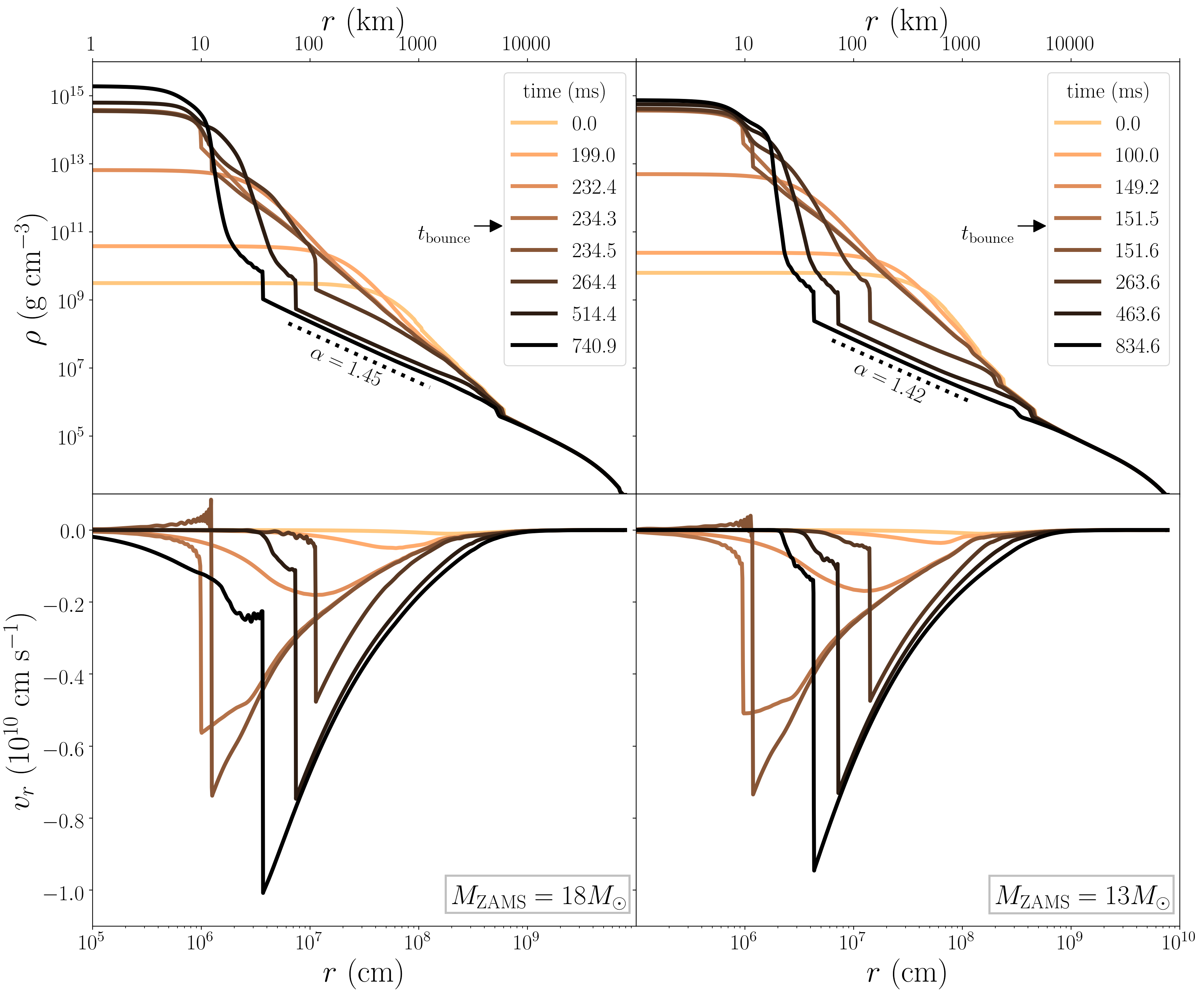}
    \caption{Evolution of the density and velocity profiles (in top and bottom panels, respectively) of the $M_\mathrm{ZAMS} = 18M_\sun$ and $13M_\sun$ models (left and right, respectively) during GR1D evolution through core collapse. The density profiles evolve similarly for both models, flattening significant after the shock stalls. In the bottom left panel, there is infall within the core (at $r < 10^{6}~\mathrm{cm}$) in the final snapshot for the $18M_\sun$ model, but no such infall in the final state of the $13M_\sun$ model. This is evidence that the core of the 18$M_\sun$ model collapses to a BH at the end of its evolution, whereas the core of the $13M_\sun$ model remains a PNS.}
    \label{fig:finalprofs_18and13}
\end{figure*}

The initial conditions for our GR1D simulations are the pre-collapse models of AD20. Each model is remapped from MESA once it has reached a maximum core infall velocity of $v_r > 1000~\mathrm{km}~\mathrm{s}^{-1}$, representing the onset of core-collapse.  The masses of the stars at this time, $M_\mathrm{pre-cc}$ are included in the third column of Table \ref{tab:modelprops}. In particular, we map the following parameters (as a function of radial coordinate) onto a new grid: enclosed mass, temperature, density, radial velocity, electron fraction, and angular velocity. We choose a grid in GR1D that is uniform ($\Delta r = 100$ m) in the inner region (up to 2 km) and logarithmically spaced outside of it, with a total of 1200 radial zones (in addition to ghost zones). The grid extends out to where the density has dropped below $\rho_\mathrm{min} = 2000~\mathrm{g~cm}^{-3}$, corresponding to a typical radius of $r(\rho = \rho_\mathrm{min}) \approx (0.8-1) \times 10^{10}~\mathrm{cm}$. This value of the minimum density does not affect the evolution of the star during collapse; the grid we use captures the overwhelming majority ($> 80\%$ by mass) of each star, and the parts of the envelope at $r \gtrsim 10^9~\mathrm{cm}$ ($\rho \lesssim 10^{5}~\mathrm{g~cm}^{-3}$) are unchanged during core-collapse (see Fig. \ref{fig:finalprofs_18and13}).

We choose a commonly-used tabulated equation of state (EOS) appropriate for hot nuclear matter from \citet{1991NuPhA.535..331L} with an incompressibility of $K_\mathrm{sat} =
220$ MeV (known as LS220). We include three species and 18 energy groups of neutrinos with tabulated opacities for a large parameter space of thermodynamic quantities generated through the open-source neutrino interaction library NuLib\footnote{\url{http://www.nulib.org/}}.

\section{Results} \label{sec:results}
\subsection{Core-collapse evolution}

The fiducial $M_\mathrm{ZAMS} = 18M_\sun$ model along with the $M_\mathrm{ZAMS} = $ 21, 40, and 45 $M_\sun$ ones all collapse to form BHs. Every model in this set follows similar evolution. On the other hand, the $M_\mathrm{ZAMS} = 13$, 25, and 35 $M_\sun$ models all fail to form BHs in the duration of the simulation and may be representative of successful neutrino-driven SN explosions. To represent these two different outcomes, we show the density and radial velocity profiles for the $M_\mathrm{ZAMS} = 18$ and 13 $M_\sun$ models at multiple stages of their evolution in Figure \ref{fig:finalprofs_18and13}.

During collapse, the outer envelope of the star ($r\gtrsim 4\times 10^{10}$ cm) is largely unaffected due to causality, while the inner region falls inward, making the core increasingly compact. Nuclear and strong forces in the dense core lead to the production of a proto-neutron star (PNS) and drive a shock outward at at time $t_\mathrm{bounce}$ (known as core-bounce). We include snapshots from just before and just after this time in Figure \ref{fig:finalprofs_18and13}.

At $t<t_\mathrm{bounce}$, infalling material steepens into a shock and there is a sharp drop in density outside the core and a correspondingly sharp negative velocity at the same radius ($r \approx 10$ km). At $t>t_\mathrm{bounce}$, the shock moves outward as neutrinos are released in the core and heat the region behind it. This is reflected in the positive velocities at the boundary of the newly formed PNS and then outward movement of the shock.

Eventually, however, the shock stalls and falls back, leading to the infall of the shock as seen in Figure \ref{fig:finalprofs_18and13} for the $M_\mathrm{ZAMS} = 18$ and 13 $M_\sun$ models. For the former, our fiducial model, core-bounce occurs at $t_\mathrm{bounce} = 234$ ms. Approximately half a second later, the PNS itself begins to collapse to form a BH, marking the end of the simulation. This is reflected in a rapid increase of the density in the inner regions and negative radial velocities within the PNS boundary at $\sim 10$ km, where the initial shock formed at core-bounce. We end our simulations at this time because GR1D can no longer evolve the metric once the central density exceeds a certain level (dependent on the choice of metric). We list the times of core bounce $t_\mathrm{bounce}$ and the final simulation time $t_f$ for all models in Table \ref{tab:modelprops}.

\subsection{Compact Remnants} \label{sec:remnants}

\begin{figure}
    \centering
    \includegraphics[width=\linewidth]{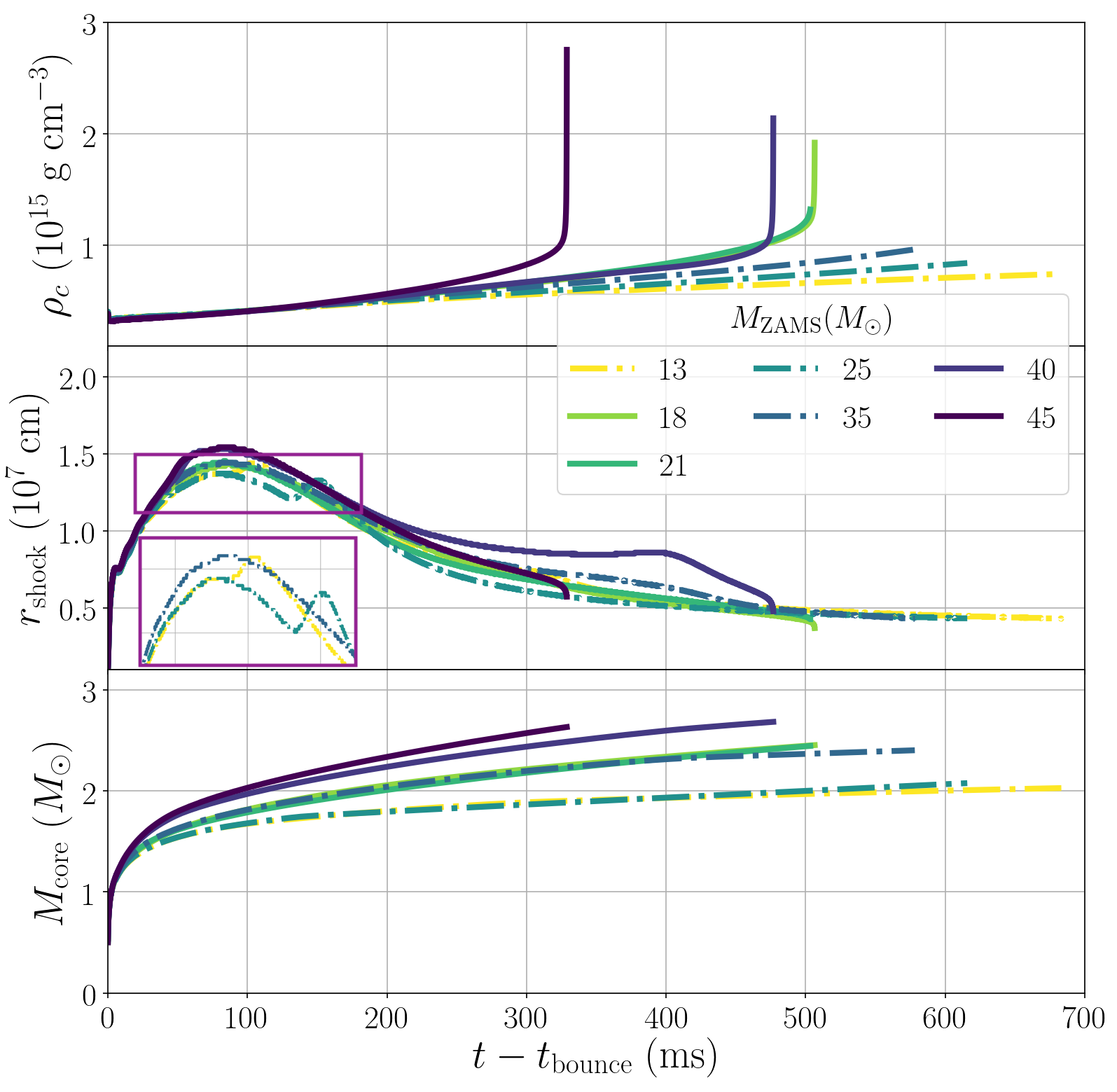}
    \caption{Evolution of several key quantities (from top to bottom: central density, shock radius, and mass of the inner core) with time after core bounce for all seven models. The shock radius $r_\mathrm{shock}$ is defined as the place where the radial velocity exceeds the sound speed and the inner core is the region within this radius, so $M_\mathrm{core} \equiv M(r<r_\mathrm{shock})$. Models that form BHs in our simulations are represented by solid lines while models that do not are shown as dash-dotted lines. The inset in the middle panel zooms in on the shock radius evolution for the three models that do not form BHs, two of which experience a brief shock revival stage.}
    \label{fig:postbounce_quants}
\end{figure}

The central density as a function of time after core-bounce is shown for each simulation in the top panel of Figure \ref{fig:postbounce_quants}. For all models, the core density grows steadily after bounce as the core continues to accrete mass. The cores of the $M_\mathrm{ZAMS} = 18M_\sun$, $21M_\sun$, $40M_\sun$, and $45M_\sun$ models all accrete enough matter to experience runaway gravitational collapse of the PNS. At this point, the central density rises exponentially as the core mass exceeds the maximum allowed NS mass, which is set by the EOS and the core angular momentum. At this stage, which indicates the onset of BH formation, the GR1D simulation ends as it cannot continue to evolve the metric to a true singularity.

In each of our seven models, a shock wave is driven outward from the core at the time of core-bounce. However, it stalls and turns back within 100 ms for all models as fallback accretion occurs. The $M_\mathrm{ZAMS} = 13$ and 25 $M_\sun$ models both experience a brief shock revival from neutrino heating but still eventually undergo a reversal and fallback accretion, as seen in the inset in the middle panel of Figure \ref{fig:postbounce_quants}. These two models also have the smallest enclosed mass within the shock radius and the lowest pre-collapse compactness parameters (see Table \ref{tab:modelprops}).

The $M_\mathrm{ZAMS}=35M_\sun$ model also fails to collapse to a BH by the end of our simulations. These outcomes suggest a critical core-compactness parameter of $\xi_{2.5} \approx 0.6$, with models that are more compact than this value collapsing to BHs in our GR1D simulations. However, it is possible that in longer or multi-dimensional simulations, some of the models that we do not see collapse in GR1D would do so. This is especially likely for the $M_\mathrm{ZAMS} = 35M_\sun$ model, which has the most massive core (as seen in Table \ref{tab:modelprops} as well as the bottom panel of Figure \ref{fig:postbounce_quants}) and the greatest pre-collapse compactness parameter (0.57) of the non-BH-forming models.

The inner core mass $M_\mathrm{core}$ is defined as the mass within the radius at which the radial velocity first exceeds the sound speed, $r_\mathrm{shock}$. While $M_\mathrm{core}$ increases throughout the duration of our simulations, the rate at which it does so decreases for all models as accretion slows and $r_\mathrm{shock}$ asymptotes. Each model forms a core of mass $2M_\sun < M_\mathrm{core} < 2.7M_\sun$ by the end of the simulation, as shown in both Table \ref{tab:modelprops} and the bottom panel of Figure \ref{fig:postbounce_quants}. In all cases, $M_\mathrm{core}$ represents a lower limit on the baryonic mass of the final remnant under the assumptions of spherical symmetry (with effective rotation) and purely neutrino-hydrodynamical evolution. The true final remnant masses may be larger, as accretion continues slowly at the end of the GR1D simulations, and may differ with the inclusion of multi-dimensional and/or magnetic effects when simulating the core-collapse evolution.

\subsection{Density Profiles}

For each of the seven simulations, we find that the density profile after core-collapse is well-fit by a distribution of the form:
\begin{equation}
    \rho(r) = \rho_0 \left(\frac{r}{r_g}\right)^{-\alpha} \left(1 - \frac{r}{R_\star}\right)^3 \label{eqn:density}
\end{equation}
where $\rho_0$ is the normalization factor satisfying $M_\star = \int_0^{R_\star} \rho(r) dV$, $r_g$ is the gravitational radius of the remnant, and $R_\star$ is the stellar radius. 

\begin{figure}
    \centering
    \includegraphics[width=\linewidth]{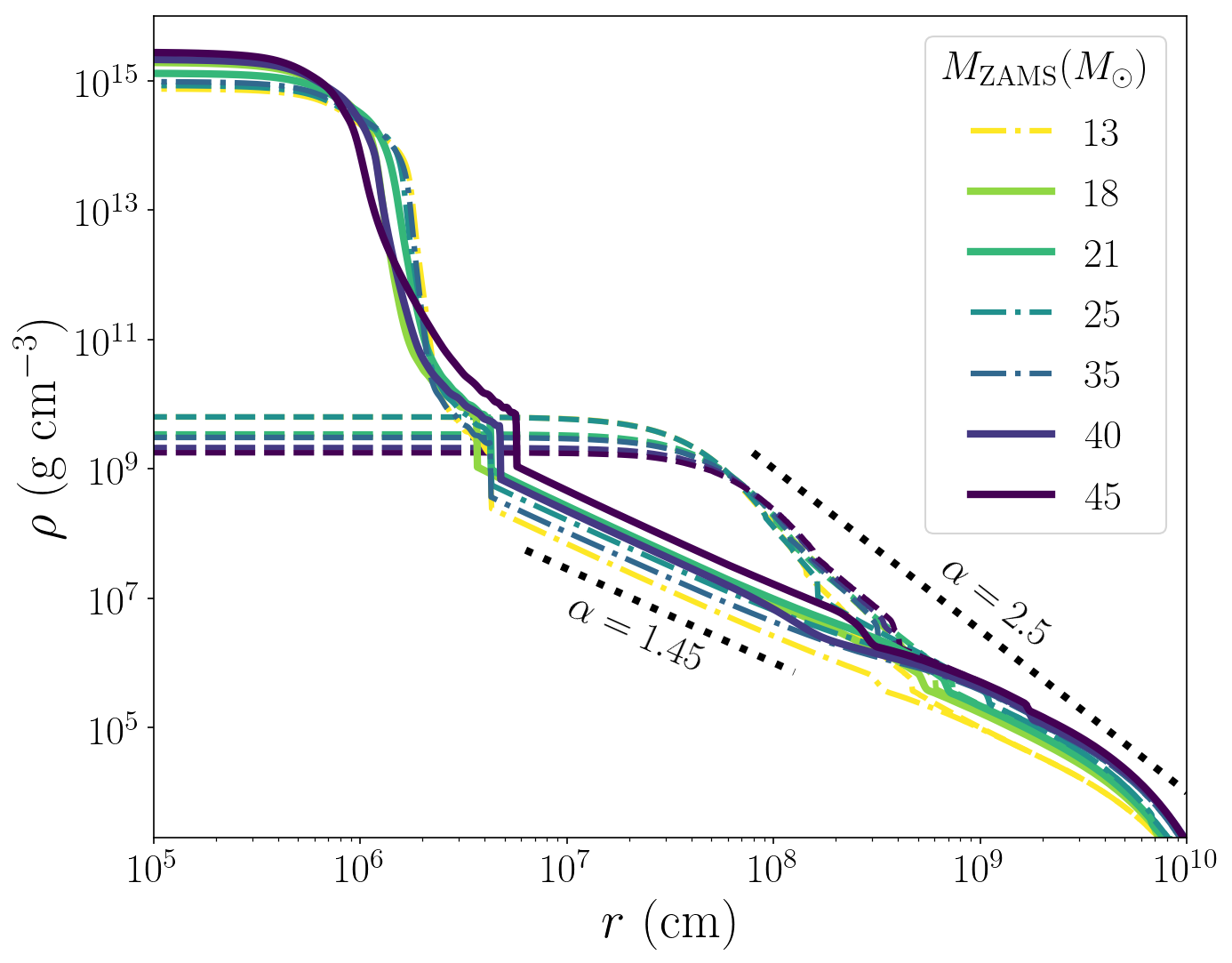}
    \caption{Initial (pre-core-collapse) and final density profiles for all seven stellar models. Initial profiles are shown as dashed lines. For the final profiles, we use solid lines for models that collapse to form BHs in our simulations and dash-dotted lines for those that do not. Representative power-law scalings are shown in dotted black lines for comparison. While all pre-collapse stellar models have steep density profiles ($\alpha \approx 2.5$), all post-collapse profiles are shallow ($\alpha \approx 1.45$). Stellar models that are initially inconsistent with lGRB observables robustly evolve to the post-collapse state necessary for jet properties consistent with observations, showing that massive stars which have undergone CHE are likely lGRB progenitors under the collapsar model.}
    \label{fig:profs_compare_all}
\end{figure}

The initial density profiles are similar across stellar masses with a best-fit power-law index of $\alpha \approx 2.5$. In all cases, the density profile becomes shallower during core-collapse, and especially after the shock stalls and turns around. The final density and pressure profiles are very smooth in the region $20~\mathrm{km} \lesssim r \lesssim 2000~\mathrm{km}$. For our fiducial collapsed stellar model at the end of the GR1D simulation, we find a best-fit value of $\alpha = 1.45$ for the density outside of the BH event horizon. We include the best-fit value of $\alpha$ for all models in the final column of Table \ref{tab:modelprops}. They vary non-monotonically with mass, though clearly trend steeper for higher values of core-compactness $\xi_{2.5}$, and range from 1.36 for the $M_\mathrm{ZAMS} = 35M_\sun$ model to 1.55 for the $M_\mathrm{ZAMS} = 40M_\sun$ model. We compare the density profiles for all seven models at the beginning and end of their GR1D simulations in Figure \ref{fig:profs_compare_all}. Outside of the cores, at radii $r>r_\mathrm{shock}$, the density distributions are very similar in shape across all models.

\section{Discussion and conclusion} \label{sec:conclusion}

Power-law density profiles with indices of $\alpha \approx 1.5$ are consistent with the simple case of free-fall acceleration outside the core. This makes it all the more meaningful that the GR1D simulations we perform yield this robust result for the final state of all models. The initial models have non-linear differences in their stellar structure, so there are complicated variations in the inputs to GR1D. These differences are mainly related to whether core carbon burning proceeds radiatively or convectively. 

Meanwhile, the physics that enters these 1D simulations is complex in that we simulate collapse in full general-relativity, with multi-group neutrino transport and realistic tabulated equations of state suitable for dense matter. Simulations that do not include neutrino transport produce dramatically different results, with post-collapse density profiles that are significantly steeper, reflecting the pre-collapse state. This suggests that neutrinos radiated from the core play an important role in shaping the thermodynamic evolution of the stellar envelope. The superficially simple outcome we find -- post-collapse stellar profiles outside the PNS or BH closely approximate those predicted by free-fall acceleration -- is thus all the more surprising.

We note that for the set of models we evolve, there is a slight positive correlation between core-compactness $\xi_{2.5}$ and the steepness of the final density profile. It remains to be seen whether this holds for a broader range of stellar evolution models rather than only those with low metallicities and rapid rotation that experience wind-driven mass loss and significant mixing, as is typical of lGRB progenitors.

\subsection{Applications for lGRBs}
The consequences of the shallow density profiles we find in our 1D core-collapse simulations of real stellar progenitor models are favorable for producing long GRBs. 3D GRMHD simulations of collapsars have followed the formation, propagation, and break-out of jets \citep{gottlieb22a,gottlieb22b}. These simulations begin with idealized initial conditions of a central BH surrounded by an effectively zero-temperature star represented by an analytic power-law density profile as in Equation (\ref{eqn:density}). Dipolar magnetic fields and fast rotational velocities are then added to the idealized stellar profile, leading to disky accretion, a build-up of magnetic flux at the BH event horizon, and the launching of a jet that propagates through the stellar envelope. \citet{gottlieb22a} varied the initial power-law index $\alpha$ and compared the physical quantities inferred from the resulting jets to observational constraints. They conclude that inner stellar density profiles with indices of $0.5 \lesssim \alpha \lesssim 1.5$ may be responsible for producing the full range of lGRB observables. In particular, such profiles are found to be necessary to produce jets with the proper luminosities, and (flat) evolution in jet power with time. On the other hand, density profiles like those of pre-collapse models 16TI \citep{woosley06} or the pre-collapse models of AD20 have steeper profiles of $\alpha \approx 2.5$. Such profiles would require an unrealistically high jet luminosity to overcome accretion and break out, $L_\mathrm{jet} \gtrsim 10^{52}~\mathrm{erg}~\mathrm{s}^{-1}$. They also produce time-evolving accretion rates, which in turn translate to evolving jet luminosities \citep{gottlieb22a}. Both of these properties would be in tension with observations of lGRBs.

Here we have shown that commonly-used pre-collapse lGRB progenitor models \citep[e.g. 16TI of][]{2006ApJ...637..914W} with density profiles that have power-law indices of $\alpha \approx 2.5$ naturally lead to post-collapse profiles with $\alpha \lesssim 1.5$. The density profiles, while flattened during collapse in the inner region of each star, are largely unchanged outside a radius of $r\gtrsim 3-5 \times 10^8~\mathrm{cm}$. Numerical studies that inject jets through an inner boundary with a radius of several thousand km are thus justified in using pre-collapse stellar profiles as the background state through which such jets propagate.

A limitation of this result is that a subset of our stellar models may successfully explode as SNe, either through the standard neutrino-driven mechanism (mainly for the $M_\mathrm{ZAMS} = 13, 35M_\sun$ models) or magneto-rotationally, when simulated in multiple dimensions with magnetic fields and rotation. The general-relativistic neutrino-hydrodynamical 1D simulations we present here do not produce SNe, so we are unable to capture the possible effects of such explosions on the resultant density profiles. Simulating the explosion process is beyond the scope of this work and outside the capabilites of GR1D, but potentially important for ensuring that stellar profiles used in lGRB simulations are fully consistent with expectations for their corresponding progenitor models, which may yield SNe. To address this limitation, we plan to perform 3D GRMHD core-collapse simulations to verify the final density profiles we obtain in the case of multi-dimensional and magnetic effects.

Another caveat is that we do not consider the possibility of a jet being driven by the PNS before BH formation. In the case of such a precursor jet, the density profile may be modified at the time of the lGRB jet launching. However, we cannot simulate this process with the techniques used here and do not require the existence of a precursor jet to produce a successful BH-driven jet. The results of \citet{gottlieb22b}, which assume a power-law density profile with $\alpha=1.5$ confirm that despite strong mixing with the star leading to baryon-loading of the jet early on, it is still able to remain intact and pierce through the envelope. As it does so, it retains its energy and relativistic velocity, generating a typical lGRB without needing to invoke a precursor jet. The stellar progenitor models which we include in this work have post-collapse density profiles akin to those chosen by \citet{gottlieb22b} and are expected to result in similarly promising jets. We plan to account for the possible signature of a precursor jet by considering non-spherically-symmetric density profiles in 3D GRMHD simulations.

In future work, we will present results from 3D GRMHD simulations that self-consistently capture jet launching, propagation, and breakout through the fiducial post-collapse stellar profile presented here with realistic thermodynamic properties \citep[][in prep.]{collapsarsII}. We emphasize that for consistency and physical accuracy, future numerical studies and theoretical models of lGRBs should consider post-collapse stellar models as the material through which lGRB jets evolve. \\

\acknowledgments
\textit{Acknowledgements:} We thank Evan O'Connor for helpful discussions and advice about using GR1D. We also thank Jim Stone, Eliot Quataert, and Elias Most for helpful discussions and the anonymous referee for insightful feedback on an earlier version of this manuscript. G.H. is supported in part by a National Science Foundation Graduate Research Fellowship. OG is supported by a CIERA Postdoctoral Fellowship. OG and AT acknowledge support by Fermi Cycle 14 Guest Investigator program 80NSSC22K0031. AT was supported by NSF
grants AST-2107839, AST-1815304, AST-1911080, OAC-2031997, and NASA grant 80NSSC18K0565. D. R. A.-D.  acknowledges support by the Stavros Niarchos Foundation (SNF) and the Hellenic Foundation for Research and Innovation (H.F.R.I.) under the 2nd Call of ``Science and Society'' Action Always strive for excellence – ``Theodoros Papazoglou'' (Project Number: 01431). The simulations presented here were performed on computational resources managed and supported by Princeton Research Computing, a consortium of groups including the Princeton Institute for Computational Science and Engineering (PICSciE) and the Office of Information Technology's High Performance Computing Center and Visualization Laboratory at Princeton University. 

\textit{Software:} GR1D \citep{oconnor10,oconnor15}, Matplotlib \citep{2007CSE.....9...90H}.

\bibliography{collapsarsI.bib}

\begin{thebibliography}{}
\expandafter\ifx\csname natexlab\endcsname\relax\def\natexlab#1{#1}\fi
\providecommand{\url}[1]{\href{#1}{#1}}

\bibitem[{{Aguilera-Dena} {et~al.}(2020){Aguilera-Dena}, {Langer},
  {Antoniadis}, \& {M{\"u}ller}}]{aguileradena20}
{Aguilera-Dena}, D.~R., {Langer}, N., {Antoniadis}, J., \& {M{\"u}ller}, B.
  2020, \apj, 901, 114

\bibitem[{{Aguilera-Dena} {et~al.}(2022){Aguilera-Dena}, {Langer},
  {Antoniadis}, {Pauli}, {Dessart}, {Vigna-G{\'o}mez}, {Gr{\"a}fener}, \&
  {Yoon}}]{2022A&A...661A..60A}
{Aguilera-Dena}, D.~R., {Langer}, N., {Antoniadis}, J., {et~al.} 2022, \aap,
  661, A60

\bibitem[{{Aguilera-Dena} {et~al.}(2018){Aguilera-Dena}, {Langer}, {Moriya}, \&
  {Schootemeijer}}]{2018ApJ...858..115A}
{Aguilera-Dena}, D.~R., {Langer}, N., {Moriya}, T.~J., \& {Schootemeijer}, A.
  2018, \apj, 858, 115

\bibitem[{{Angus} {et~al.}(2016){Angus}, {Levan}, {Perley}, {Tanvir}, {Lyman},
  {Stanway}, \& {Fruchter}}]{2016MNRAS.458...84A}
{Angus}, C.~R., {Levan}, A.~J., {Perley}, D.~A., {et~al.} 2016, \mnras, 458, 84

\bibitem[{{Barnes} {et~al.}(2018){Barnes}, {Duffell}, {Liu}, {Modjaz},
  {Bianco}, {Kasen}, \& {MacFadyen}}]{2018ApJ...860...38B}
{Barnes}, J., {Duffell}, P.~C., {Liu}, Y., {et~al.} 2018, \apj, 860, 38

\bibitem[{{Blandford} \& {Znajek}(1977)}]{blandford77}
{Blandford}, R.~D., \& {Znajek}, R.~L. 1977, \mnras, 179, 433

\bibitem[{{Bloom} {et~al.}(2002){Bloom}, {Kulkarni}, \& {Djorgovski}}]{bloom02}
{Bloom}, J.~S., {Kulkarni}, S.~R., \& {Djorgovski}, S.~G. 2002, \aj, 123, 1111

\bibitem[{{Burrows} {et~al.}(2007){Burrows}, {Dessart}, {Livne}, {Ott}, \&
  {Murphy}}]{burrows07}
{Burrows}, A., {Dessart}, L., {Livne}, E., {Ott}, C.~D., \& {Murphy}, J. 2007,
  \apj, 664, 416

\bibitem[{{Dessart} {et~al.}(2017){Dessart}, {Hillier}, {Yoon}, {Waldman}, \&
  {Livne}}]{2017A&A...603A..51D}
{Dessart}, L., {Hillier}, D.~J., {Yoon}, S.-C., {Waldman}, R., \& {Livne}, E.
  2017, \aap, 603, A51

\bibitem[{{Eisenberg} {et~al.}(2022){Eisenberg}, {Gottlieb}, \&
  {Nakar}}]{2022MNRAS.517..582E}
{Eisenberg}, M., {Gottlieb}, O., \& {Nakar}, E. 2022, \mnras, 517, 582

\bibitem[{{Ertl} {et~al.}(2016){Ertl}, {Janka}, {Woosley}, {Sukhbold}, \&
  {Ugliano}}]{ertl2016}
{Ertl}, T., {Janka}, H.~T., {Woosley}, S.~E., {Sukhbold}, T., \& {Ugliano}, M.
  2016, \apj, 818, 124

\bibitem[{{Gal-Yam}(2012)}]{2012Sci...337..927G}
{Gal-Yam}, A. 2012, Science, 337, 927

\bibitem[{{Gal-Yam}(2019)}]{2019ApJ...882..102G}
---. 2019, \apj, 882, 102

\bibitem[{{Gal-Yam} {et~al.}(2014){Gal-Yam}, {Arcavi}, {Ofek}, {Ben-Ami},
  {Cenko}, {Kasliwal}, {Cao}, {Yaron}, {Tal}, {Silverman}, {Horesh}, {De Cia},
  {Taddia}, {Sollerman}, {Perley}, {Vreeswijk}, {Kulkarni}, {Nugent},
  {Filippenko}, \& {Wheeler}}]{2014Natur.509..471G}
{Gal-Yam}, A., {Arcavi}, I., {Ofek}, E.~O., {et~al.} 2014, \nat, 509, 471

\bibitem[{{Galama} {et~al.}(1998){Galama}, {Vreeswijk}, {van Paradijs},
  {Kouveliotou}, {Augusteijn}, {B{\"o}hnhardt}, {Brewer}, {Doublier},
  {Gonzalez}, {Leibundgut}, {Lidman}, {Hainaut}, {Patat}, {Heise}, {in't Zand},
  {Hurley}, {Groot}, {Strom}, {Mazzali}, {Iwamoto}, {Nomoto}, {Umeda},
  {Nakamura}, {Young}, {Suzuki}, {Shigeyama}, {Koshut}, {Kippen}, {Robinson},
  {de Wildt}, {Wijers}, {Tanvir}, {Greiner}, {Pian}, {Palazzi}, {Frontera},
  {Masetti}, {Nicastro}, {Feroci}, {Costa}, {Piro}, {Peterson}, {Tinney},
  {Boyle}, {Cannon}, {Stathakis}, {Sadler}, {Begam}, \& {Ianna}}]{galama98}
{Galama}, T.~J., {Vreeswijk}, P.~M., {van Paradijs}, J., {et~al.} 1998, \nat,
  395, 670

\bibitem[{{Gottlieb} {et~al.}(2022{\natexlab{a}}){Gottlieb}, {Lalakos},
  {Bromberg}, {Liska}, \& {Tchekhovskoy}}]{gottlieb22a}
{Gottlieb}, O., {Lalakos}, A., {Bromberg}, O., {Liska}, M., \& {Tchekhovskoy},
  A. 2022{\natexlab{a}}, \mnras, 510, 4962

\bibitem[{{Gottlieb} {et~al.}(2022{\natexlab{b}}){Gottlieb}, {Liska},
  {Tchekhovskoy}, {Bromberg}, {Lalakos}, {Giannios}, \&
  {M{\"o}sta}}]{gottlieb22b}
{Gottlieb}, O., {Liska}, M., {Tchekhovskoy}, A., {et~al.} 2022{\natexlab{b}},
  \apjl, 933, L9

\bibitem[{{Grevesse} {et~al.}(1996){Grevesse}, {Noels}, \&
  {Sauval}}]{grevesse96}
{Grevesse}, N., {Noels}, A., \& {Sauval}, A.~J. 1996, in Astronomical Society
  of the Pacific Conference Series, Vol.~99, Cosmic Abundances, ed. S.~S.
  {Holt} \& G.~{Sonneborn}, 117

\bibitem[{{Halevi} {et~al.}(2023){Halevi}, {Gottlieb}, {M\"{o}sta},
  {Tchekhovskoy}, {Aguilera-Dena}, \& {Wu}}]{collapsarsII}
{Halevi}, G., {Gottlieb}, O., {M\"{o}sta}, P., {et~al.} 2023, expected
  submission to \apj

\bibitem[{{Halevi} \& {M{\"o}sta}(2018)}]{halevi18}
{Halevi}, G., \& {M{\"o}sta}, P. 2018, \mnras, 477, 2366

\bibitem[{{Hunter}(2007)}]{2007CSE.....9...90H}
{Hunter}, J.~D. 2007, Computing in Science and Engineering, 9, 90

\bibitem[{{Kasen} \& {Bildsten}(2010)}]{2010ApJ...717..245K}
{Kasen}, D., \& {Bildsten}, L. 2010, \apj, 717, 245

\bibitem[{{Komissarov} \& {Barkov}(2009)}]{komissarov09}
{Komissarov}, S.~S., \& {Barkov}, M.~V. 2009, \mnras, 397, 1153

\bibitem[{{Lattimer} \& {Swesty}(1991)}]{1991NuPhA.535..331L}
{Lattimer}, J.~M., \& {Swesty}, D.~F. 1991, \nphysa, 535, 331

\bibitem[{{Lazzati} {et~al.}(2012){Lazzati}, {Morsony}, {Blackwell}, \&
  {Begelman}}]{lazzati12}
{Lazzati}, D., {Morsony}, B.~J., {Blackwell}, C.~H., \& {Begelman}, M.~C. 2012,
  \apj, 750, 68

\bibitem[{{Lazzati} {et~al.}(2015){Lazzati}, {Morsony}, \&
  {L{\'o}pez-C{\'a}mara}}]{2015JHEAp...7...17L}
{Lazzati}, D., {Morsony}, B.~J., \& {L{\'o}pez-C{\'a}mara}, D. 2015, Journal of
  High Energy Astrophysics, 7, 17

\bibitem[{{L{\'o}pez-C{\'a}mara} {et~al.}(2016){L{\'o}pez-C{\'a}mara},
  {Lazzati}, \& {Morsony}}]{2016ApJ...826..180L}
{L{\'o}pez-C{\'a}mara}, D., {Lazzati}, D., \& {Morsony}, B.~J. 2016, \apj, 826,
  180

\bibitem[{{L{\'o}pez-C{\'a}mara} {et~al.}(2013){L{\'o}pez-C{\'a}mara},
  {Morsony}, {Begelman}, \& {Lazzati}}]{2013ApJ...767...19L}
{L{\'o}pez-C{\'a}mara}, D., {Morsony}, B.~J., {Begelman}, M.~C., \& {Lazzati},
  D. 2013, \apj, 767, 19

\bibitem[{{Lunnan} {et~al.}(2014){Lunnan}, {Chornock}, {Berger}, {Laskar},
  {Fong}, {Rest}, {Sanders}, {Challis}, {Drout}, {Foley}, {Huber}, {Kirshner},
  {Leibler}, {Marion}, {McCrum}, {Milisavljevic}, {Narayan}, {Scolnic},
  {Smartt}, {Smith}, {Soderberg}, {Tonry}, {Burgett}, {Chambers}, {Flewelling},
  {Hodapp}, {Kaiser}, {Magnier}, {Price}, \& {Wainscoat}}]{2014ApJ...787..138L}
{Lunnan}, R., {Chornock}, R., {Berger}, E., {et~al.} 2014, \apj, 787, 138

\bibitem[{{MacFadyen} \& {Woosley}(1999)}]{macfadyen99}
{MacFadyen}, A.~I., \& {Woosley}, S.~E. 1999, \apj, 524, 262

\bibitem[{{Margalit} {et~al.}(2018){Margalit}, {Metzger}, {Thompson},
  {Nicholl}, \& {Sukhbold}}]{2018MNRAS.475.2659M}
{Margalit}, B., {Metzger}, B.~D., {Thompson}, T.~A., {Nicholl}, M., \&
  {Sukhbold}, T. 2018, \mnras, 475, 2659

\bibitem[{{Mazzali} {et~al.}(2016){Mazzali}, {Sullivan}, {Pian}, {Greiner}, \&
  {Kann}}]{2016MNRAS.458.3455M}
{Mazzali}, P.~A., {Sullivan}, M., {Pian}, E., {Greiner}, J., \& {Kann}, D.~A.
  2016, \mnras, 458, 3455

\bibitem[{{Metzger} {et~al.}(2011){Metzger}, {Giannios}, {Thompson},
  {Bucciantini}, \& {Quataert}}]{protomagnetar_GRB}
{Metzger}, B.~D., {Giannios}, D., {Thompson}, T.~A., {Bucciantini}, N., \&
  {Quataert}, E. 2011, \mnras, 413, 2031

\bibitem[{{Metzger} {et~al.}(2015){Metzger}, {Margalit}, {Kasen}, \&
  {Quataert}}]{2015MNRAS.454.3311M}
{Metzger}, B.~D., {Margalit}, B., {Kasen}, D., \& {Quataert}, E. 2015, \mnras,
  454, 3311

\bibitem[{{Modjaz} {et~al.}(2016){Modjaz}, {Liu}, {Bianco}, \&
  {Graur}}]{modjaz16}
{Modjaz}, M., {Liu}, Y.~Q., {Bianco}, F.~B., \& {Graur}, O. 2016, \apj, 832,
  108

\bibitem[{{Modjaz} {et~al.}(2006){Modjaz}, {Stanek}, {Garnavich}, {Berlind},
  {Blondin}, {Brown}, {Calkins}, {Challis}, {Diamond-Stanic}, {Hao}, {Hicken},
  {Kirshner}, \& {Prieto}}]{modjaz06}
{Modjaz}, M., {Stanek}, K.~Z., {Garnavich}, P.~M., {et~al.} 2006, \apjl, 645,
  L21

\bibitem[{{Moriya} {et~al.}(2018){Moriya}, {Sorokina}, \&
  {Chevalier}}]{2018SSRv..214...59M}
{Moriya}, T.~J., {Sorokina}, E.~I., \& {Chevalier}, R.~A. 2018, \ssr, 214, 59

\bibitem[{{Morsony} {et~al.}(2007){Morsony}, {Lazzati}, \&
  {Begelman}}]{2007ApJ...665..569M}
{Morsony}, B.~J., {Lazzati}, D., \& {Begelman}, M.~C. 2007, \apj, 665, 569

\bibitem[{{M{\"o}sta} {et~al.}(2014){M{\"o}sta}, {Richers}, {Ott}, {Haas},
  {Piro}, {Boydstun}, {Abdikamalov}, {Reisswig}, \& {Schnetter}}]{mosta14}
{M{\"o}sta}, P., {Richers}, S., {Ott}, C.~D., {et~al.} 2014, \apjl, 785, L29

\bibitem[{{M{\"u}ller}(2016)}]{2016PASA...33...48M}
{M{\"u}ller}, B. 2016, \pasa, 33, e048

\bibitem[{{M{\"u}ller} {et~al.}(2016){M{\"u}ller}, {Heger}, {Liptai}, \&
  {Cameron}}]{muller2016}
{M{\"u}ller}, B., {Heger}, A., {Liptai}, D., \& {Cameron}, J.~B. 2016, \mnras,
  460, 742

\bibitem[{{Nicholl} {et~al.}(2017){Nicholl}, {Guillochon}, \&
  {Berger}}]{2017ApJ...850...55N}
{Nicholl}, M., {Guillochon}, J., \& {Berger}, E. 2017, \apj, 850, 55

\bibitem[{{Obergaulinger} \& {Aloy}(2020)}]{2020MNRAS.492.4613O}
{Obergaulinger}, M., \& {Aloy}, M.~{\'A}. 2020, \mnras, 492, 4613

\bibitem[{{O'Connor}(2015)}]{oconnor15}
{O'Connor}, E. 2015, \apjs, 219, 24

\bibitem[{{O'Connor} \& {Ott}(2010)}]{oconnor10}
{O'Connor}, E., \& {Ott}, C.~D. 2010, Classical and Quantum Gravity, 27, 114103

\bibitem[{{O'Connor} \& {Ott}(2011)}]{2011ApJ...730...70O}
---. 2011, \apj, 730, 70

\bibitem[{{O'Connor} {et~al.}(2018){O'Connor}, {Bollig}, {Burrows}, {Couch},
  {Fischer}, {Janka}, {Kotake}, {Lentz}, {Liebend{\"o}rfer}, {Messer},
  {Mezzacappa}, {Takiwaki}, \& {Vartanyan}}]{2018JPhG...45j4001O}
{O'Connor}, E., {Bollig}, R., {Burrows}, A., {et~al.} 2018, Journal of Physics
  G Nuclear Physics, 45, 104001

\bibitem[{{O'Connor} \& {Couch}(2018)}]{2018ApJ...865...81O}
{O'Connor}, E.~P., \& {Couch}, S.~M. 2018, \apj, 865, 81

\bibitem[{{Paxton} {et~al.}(2011){Paxton}, {Bildsten}, {Dotter}, {Herwig},
  {Lesaffre}, \& {Timmes}}]{Paxton2011}
{Paxton}, B., {Bildsten}, L., {Dotter}, A., {et~al.} 2011, \apjs, 192, 3

\bibitem[{{Paxton} {et~al.}(2013){Paxton}, {Cantiello}, {Arras}, {Bildsten},
  {Brown}, {Dotter}, {Mankovich}, {Montgomery}, {Stello}, {Timmes}, \&
  {Townsend}}]{Paxton2013}
{Paxton}, B., {Cantiello}, M., {Arras}, P., {et~al.} 2013, \apjs, 208, 4

\bibitem[{{Paxton} {et~al.}(2015){Paxton}, {Marchant}, {Schwab}, {Bauer},
  {Bildsten}, {Cantiello}, {Dessart}, {Farmer}, {Hu}, {Langer}, {Townsend},
  {Townsley}, \& {Timmes}}]{Paxton2015}
{Paxton}, B., {Marchant}, P., {Schwab}, J., {et~al.} 2015, \apjs, 220, 15

\bibitem[{{Paxton} {et~al.}(2018){Paxton}, {Schwab}, {Bauer}, {Bildsten},
  {Blinnikov}, {Duffell}, {Farmer}, {Goldberg}, {Marchant}, {Sorokina},
  {Thoul}, {Townsend}, \& {Timmes}}]{Paxton2018}
{Paxton}, B., {Schwab}, J., {Bauer}, E.~B., {et~al.} 2018, \apjs, 234, 34

\bibitem[{{Shankar} {et~al.}(2021){Shankar}, {M{\"o}sta}, {Barnes}, {Duffell},
  \& {Kasen}}]{protomagnetar_IcBL}
{Shankar}, S., {M{\"o}sta}, P., {Barnes}, J., {Duffell}, P.~C., \& {Kasen}, D.
  2021, \mnras, 508, 5390

\bibitem[{{Shenar} {et~al.}(2020){Shenar}, {Gilkis}, {Vink}, {Sana}, \&
  {Sander}}]{2020A&A...634A..79S}
{Shenar}, T., {Gilkis}, A., {Vink}, J.~S., {Sana}, H., \& {Sander}, A.~A.~C.
  2020, \aap, 634, A79

\bibitem[{{Song} \& {Liu}(2023)}]{2023arXiv230105401S}
{Song}, C.-Y., \& {Liu}, T. 2023, arXiv e-prints, arXiv:2301.05401

\bibitem[{{Sukhbold} \& {Woosley}(2014)}]{sukhbold2014}
{Sukhbold}, T., \& {Woosley}, S.~E. 2014, \apj, 783, 10

\bibitem[{{Sz{\'e}csi} {et~al.}(2015){Sz{\'e}csi}, {Langer}, {Yoon}, {Sanyal},
  {de Mink}, {Evans}, \& {Dermine}}]{2015A&A...581A..15S}
{Sz{\'e}csi}, D., {Langer}, N., {Yoon}, S.-C., {et~al.} 2015, \aap, 581, A15

\bibitem[{{Woosley}(1993)}]{woosley93}
{Woosley}, S.~E. 1993, \apj, 405, 273

\bibitem[{{Woosley}(2010)}]{2010ApJ...719L.204W}
---. 2010, \apjl, 719, L204

\bibitem[{{Woosley} \& {Heger}(2006{\natexlab{a}})}]{2006ApJ...637..914W}
{Woosley}, S.~E., \& {Heger}, A. 2006{\natexlab{a}}, \apj, 637, 914

\bibitem[{{Woosley} \& {Heger}(2006{\natexlab{b}})}]{woosley06}
---. 2006{\natexlab{b}}, \apj, 637, 914

\bibitem[{{Xie} \& {MacFadyen}(2019)}]{2019ApJ...880..135X}
{Xie}, X., \& {MacFadyen}, A. 2019, \apj, 880, 135

\end{thebibliography}

\end{document}